\documentclass[12pt,a4paper]{article}

\usepackage[english]{babel}

\usepackage{indentfirst}
\usepackage{amsmath}
\usepackage{amssymb}
\usepackage{soul}
\usepackage[pdftex]{graphicx,color}

\definecolor{green}{rgb}{0.0,0.6,0.0}

\usepackage{geometry}
\begin{document}

\begin{flushright}
IPPP/19/25
\\
%\today
\end{flushright}
\bigskip

%\hfill \today

\begin{center}
\textbf{\LARGE \boldmath 
$\Delta A_{CP}$ within the Standard Model and beyond}

\vspace*{1.6cm}

{\large Mikael Chala, Alexander Lenz, Aleksey V. Rusov and Jakub Scholtz}

\vspace*{0.4cm}

\textsl{%
Institute for Particle Physics Phenomenology, Durham University, \\
DH1 3LE Durham, United Kingdom \\[5mm]
}
\vspace*{0.8cm}

\textbf{\Large Abstract}\\[10pt]
\parbox[t]{0.9\textwidth}{In light of the recent LHCb observation of CP violation in the charm sector, we  review
standard model (SM) predictions in the charm sector and in particular for $\Delta A_{CP}$. We get as an upper bound in the 
SM $| \Delta A_{CP} ^{\rm SM}|  \leq 3.6 \times 10^{-4}$, which can be compared to the measurement of 
 $\Delta A_{CP} ^{\rm LHCb2019} = (-15.4 \pm 2.9)  \times 10^{-4}$. We discuss resolving this tension within an extension of the SM that includes a flavour violating $Z'$ that couples only to $\bar{s}s$ and $\bar{c}u$. We show that for masses below 80 GeV  and flavour violating coupling of  the order of $10^{-4}$, this model can successfully resolve the tension and avoid constraints from dijet searches, $D^0-\overline{D}^0$ mixing and measurements of the $Z$ width.
 
 }

\end{center}

\vspace*{3cm}
\newpage

\section{Introduction}
CP violation has so far been firmly established in the down-quark sector, while similar effects in the charm-quark sector 
were expected to be tiny. In 2011 the LHCb Collaboration reported \cite{Aaij:2011in} the first evidence for such an effect 
in the quantity
\begin{eqnarray}
\Delta A_{CP} & = & A_{CP} (K^-  K^+ )  -  A_{CP} (\pi^- \pi^+ ) \, ,
\end{eqnarray}
where the time dependent asymmetry into a final state $f$ is given by
\begin{eqnarray}
 A_{CP} (f,t ) & = & \frac{\Gamma (D^0(t) \to f) - \Gamma (\overline{D}^0(t) \to f)} {\Gamma (D^0(t) \to f) + \Gamma (\overline{D}^0(t) \to f)}  \, .
\end{eqnarray}
This asymmetry can be further decomposed into a direct CP asymmetry and a mixing induced CP asymmetry:
\begin{eqnarray}
 A_{CP} (f,t ) & = & a_{CP}^{\rm dir}(f) +  \frac{t}{\tau (D^0)} a_{CP}^{\rm ind} \, ,
 \end{eqnarray}
 where $\tau$ is the lifetime of the neutral $D$ meson.
The flavour of the initial state ($D^0$ or $\overline{D}^0$) can either be tagged by identifying the charge of the pion in the decay
$D^{+*} \to D^0 + \pi^+$ (pion tag) or by identifying the muon in the decay $B \to D^0 \mu^-X$ (muon tag).
Originally, the large effect in $\Delta A_{CP}$ was confirmed by CDF \cite{Collaboration:2012qw} and Belle \cite{Ko:2012px}. Later on, the effect was not seen 
in an LHCb analysis based on muon tag \cite{Aaij:2013bra,Aaij:2014gsa} and it also disappeared largely in the pion tag analysis \cite{Aaij:2016cfh}.
At that point in time, the theoretical interpretation of a large direct CP violation was also rather inconclusive, see \textit{e.g.} Ref.~\cite{Lenz:2013pwa}
and it was not clear whether a large value of $\Delta A_{CP}$ could still be due to 
underestimated non-perturbative effects 
(see \textit{e.g.} Refs.~\cite{Golden:1989qx,Pirtskhalava:2011va,Bhattacharya:2012ah,Brod:2011re,Feldmann:2012js,Brod:2012ud,Franco:2012ck,Cheng:2012wr,Cheng:2012xb})
or whether this was already a clear indication of new physics
(see \textit{e.g.} Refs.~\cite{Isidori:2011qw,Giudice:2012qq,Altmannshofer:2012ur,Hiller:2012wf}). 
At Moriond 2019, the LHCb collaboration presented new measurements \cite{Aaij:2019kcg} and the combined
value is currently
\begin{equation}
\Delta A_{CP} ^{\rm E
xp.}= (-15.4 \pm 2.9) \times 10^{-4} \, ,
\label{DeltaACP_EXP}
\end{equation}
being $5.3$ standard deviations away from zero and originating mostly from direct CP violation. See Tab.~\ref{tab:time} for a summary of experimental results and their references.
  
\begin{table}[t]
\begin{center}
\begin{tabular}{|l||c|c|c|c|}
\hline
\mbox{Experiment}   &  $\Delta A_{CP} \times 10^4$  & \mbox{Tag}  & \mbox{arXiv} & \mbox{Reference}
\\
\hline
\hline
 \mbox{BaBar} & $+24 \pm 62 \pm  26$ & \mbox{pion} & 0709.2715      & \cite{Aubert:2007if}
\\
\hline
\mbox{LHCb} & $-  82 \pm 21 \pm  11$ & \mbox{pion} &1112.0938      & \cite{Aaij:2011in}
\\
\hline
\mbox{CDF} & $-  62 \pm 21 \pm  10$ & \mbox{pion} &1207.2158      & \cite{Collaboration:2012qw}
\\
\hline
\mbox{Belle} & $-  87 \pm 41 \pm  6$ & \mbox{pion} &1212.1975      & \cite{Ko:2012px}
\\
\hline
\mbox{LHCb} & $+49 \pm 30 \pm  14$ & \mbox{muon} &1303.2614      & \cite{Aaij:2013bra}
\\
\hline
\mbox{LHCb} & $+14 \pm 16 \pm  8$ & \mbox{muon} &1405.2797     & \cite{Aaij:2014gsa}
\\
\hline
\mbox{LHCb} & $-10 \pm 8 \pm  3$ & \mbox{pion} & 1602.03160  & \cite{Aaij:2016cfh}
\\
\hline
\mbox{LHCb} & $-18.2 \pm 3.2 \pm  0.9$ & \mbox{pion} &1903.08726  & \cite{Aaij:2019kcg}
\\
\hline
\mbox{LHCb} & $-9 \pm 8 \pm  5$ & \mbox{muon} &  1903.08726  & \cite{Aaij:2019kcg}
\\
\hline
\end{tabular}
\caption{\it Time evolution of CP violation in the charm sector.}\label{tab:time}
\end{center}
\end{table}

\section{Standard Model predictions in the charm sector}
Reliable theory predictions in the charm sector seem to be notoriously difficult. 
Sometimes the famous $\Delta I = 1/2$ rule in the Kaon sector is given as a motivation 
for very large hadronic effects in the charm sector (see \textit{e.g.}
Refs.~\cite{Golden:1989qx,Pirtskhalava:2011va,Bhattacharya:2012ah,Brod:2012ud}).
This argument has, however, several flaws. First, the $D \to \pi \pi$ data show no enhancement 
of the $\Delta I = 1/2$ over the $\Delta I = 3/2$ amplitude, see Ref.~\cite{Franco:2012ck}. 
Next, modern theoretical explanations of the $\Delta I = 1/2$ rule in the Kaon sector, 
based on lattice calculations \cite{Boyle:2012ys,Bai:2015nea} see no enhancement of penguin
contributions (this would be required for a large value of $\Delta A_{CP}$), 
but a severe cancellation of tree level contributions (see \cite{Atwood:2012ac} for a similar comment).
Finally, the extrapolation of results
from the strange sector to the charm seems to lack a theoretical foundation.

We will briefly review the
prime example where the SM seems to be orders of magnitudes off: charm mixing. On the other
hand, we found recently that charm lifetimes can be unexpectedly well described within the SM. Finally, 
we show that the seemingly huge discrepancy in charm mixing could actually be due to small (as low as $20\%$)
unknown non-perturbative effects.
\subsection{Charm mixing}
Diagonalisation of the $2\times 2$ matrix describing the mixing of the neutral $D$ mesons
gives the same eigenvalue equations as in the neutral $B$ systems:
\begin{eqnarray}
\Delta M_D^2 - \frac14 \Delta \Gamma_D^2 
=  
4 \left| M_{12}^D \right|^2 - \left| \Gamma_{12}^D \right|^2 \, ,
&&
\Delta M_D \Delta \Gamma_D 
 =  
4 \left| M_{12}^D \right| \left| \Gamma_{12}^D \right| \cos (\phi_{12}^D) \, ,
\end{eqnarray}
where $\Delta M_D$ is the mass difference and $\Delta \Gamma_D$ is the decay rate difference of the mass 
eigenstates of the neutral $D$ mesons. The box diagrams giving rise to $D$ mixing can have internal
$d$, $s$ and $b$ quarks --- compared to $u$, $c$, $t$ in the $B$ sector.  $M_{12}^D$ denotes the 
dispersive part of the box diagram, $\Gamma_{12}^D$ the absorptive part and the relative phase 
of the two is given by $\phi_{12}^D = - \arg (-M_{12}^D / \Gamma_{12}^D)$. Unlike in the $B$ 
system, where $|\Gamma_{12}/M_{12}| \ll 1$ holds, the expressions for  $\Delta M_D$ and 
$\Delta \Gamma_D$ in terms of $M_{12}^D$ and $\Gamma_{12}^D$ can not be simplified,  
and both $M_{12}^D$ and $\Gamma_{12}^D$ have to be known in order to compute $\Delta M_D$ or $\Delta \Gamma_D$. 
On the other hand, it is well-known that bounds like $\Delta \Gamma_D \leq 2 |\Gamma_{12}^D |$ 
hold \cite{Nierste:2009wg,Jubb:2016mvq}.
The experimental measurements (webupdate of Ref.~\cite{Amhis:2016xyh})
of the mass and decay rate differences 
yield very small values
\begin{eqnarray}
x \equiv \Delta M_D / \Gamma_D = \left( 0.39^{+0.11}_{-0.12} \right)   \% \, ,
&&
y \equiv \Delta \Gamma_D / (2 \Gamma_D) = \left( 0.651^{+0.063}_{-0.069} \right) \%  \, ,
\end{eqnarray}
where $\Gamma_D$ denotes the total decay rate of the neutral $D$ mesons. While
a decay rate difference in the neutral $D$ system is by now firmly established, the possibility of 
having a vanishing mass difference is still not excluded --- the strongest evidence is currently coming from the measurement
reported in Ref.~\cite{Aaij:2019jot} and the future experimental sensitivity for $x$ and $y$
will be at the order of $0.005 \%$ \cite{Cerri:2018ypt}.

The on-shell contribution $\Gamma_{12}^D$ can be expressed in terms of  box diagrams differing in the 
internal quarks --- $(s \bar{s})$, $(s \bar{d})$, $(d \bar{s})$ and $(d \bar{d})$.
Using the unitarity of the CKM matrix \cite{Cabibbo:1963yz,Kobayashi:1973fv}, namely $\lambda_d + \lambda_s +\lambda_b = 0$ with  $\lambda_x = V_{cx} V^*_{ux}$, one gets
\begin{eqnarray}
\Gamma_{12}^D & = & -   \lambda_s^2 \left(\Gamma_{ss}^D - 2 \Gamma_{sd}^D + \Gamma_{dd}^D\right)
                    + 2 \lambda_s \lambda_b \left(\Gamma_{sd}^D - \Gamma_{dd}^D \right)
                    -   \lambda_b^2 \Gamma_{dd}^D \, .
\label{Gamma12D}
\end{eqnarray}
Eq.~(\ref{Gamma12D}) shows a very pronounced CKM hierarchy: 
expressed in terms of the Wolfenstein parameter \cite{Wolfenstein:1983yz} $\lambda \approx 0.225$ 
(webupdate of Refs.~\cite{Charles:2004jd,Bona:2006ah}) one has $\lambda_s \propto \lambda$ 
and $\lambda_b \propto \lambda^5$. 
In the exact $SU(3)_F$ limit, $\Gamma_{ss}^D = \Gamma_{sd}^D = \Gamma_{dd}^D$ holds and 
the first two terms of the r.h.s. of Eq.~(\ref{Gamma12D}) vanish and only the tiny contribution from 
the third term survives. 
The determination of $M_{12}^D$ involves, in addition, box diagrams with internal $b$ 
quarks and in contrast to $\Gamma_{12}^D$, the dispersive part of the diagrams has 
to be determined. Denoting the dispersive part of a box diagram with internal $i$ and $j$ 
quarks by $M_{ij}^D$ and using CKM unitarity again one gets the following structure:
\begin{eqnarray}
\label{M12D}
M_{12}^D \! & \! = \! &  \! \! \! \lambda_s^2          \left[ M_{ss}^D \! - 2 M_{sd}^D + M_{dd}^D \right]\\\nonumber
              & & \! + 2 \lambda_s \lambda_b  \left[ M_{bs}^D \! -   M_{bd}^D \! - M_{sd}^D + M_{dd}^D \right]
              \!+   \lambda_b^2          \left[ M_{bb}^D \! - 2 M_{bd}^D + M_{dd}^D \right]\, .
\end{eqnarray} 
In the case of neutral $B$ mesons, the third term (replacing $b,s,d \to t,c,u$) is clearly dominant, while in the case 
of $D$ mesons the extreme CKM suppression of $\lambda_b$ might be compensated by a less pronounced 
GIM cancellation \cite{Glashow:1970gm} and in the end all three contributions of 
Eq.~(\ref{M12D}) could have a similar size.

For the theoretical determination of $M_{12}^D$ and $\Gamma_{12}^D$, one can use a 
quark-level (inclusive) or a hadron-level (exclusive) description. 
The inclusive approach for $\Gamma_{12}^D$ is based on the heavy quark expansion (HQE) 
\cite{Khoze:1983yp,Shifman:1984wx,Bigi:1991ir,Bigi:1992su,Blok:1992hw,Blok:1992he,Beneke:1998sy} and 
works very well for the $B$ system \cite{Lenz:2014jha,Artuso:2015swg,Kirk:2017juj}.
Applying the HQE (the relevant non-perturbative matrix elements of dimension six operators have been
determined in Refs.~\cite{Carrasco:2014uya,Carrasco:2015pra,Bazavov:2017weg,Kirk:2017juj}) to 
a single diagram contributing to $\Gamma_{12}^D$ -- \textit{e.g.} only internal
$s \bar{s}$ quark -- one gets five times the experimental value of $y$ \cite{Lenz:2016fcv}.
Applying the HQE to the whole expression of Eq.~(\ref{Gamma12D}) leads to  an extremely severe GIM cancellation and the 
overall result lies about four orders of magnitude below the experiment! Given that the HQE succeeds in the $B$ system\footnote{In the $B$ system the 
expansion parameter is only a factor of three smaller.} and for $D$ meson lifetimes, which we will discuss below, 
it is unlikely that the HQE fails by four orders of magnitude in charm mixing. Instead, the problem seems
to be rooted in severe GIM cancellations.

The exclusive approach \cite{Falk:2001hx,Falk:2004wg,Cheng:2010rv,Gershon:2015xra,Jiang:2017zwr} 
aims to determine $M_{12}^D$ and  $\Gamma_{12}^D$ at the hadron level. A potential starting point are the expressions
\begin{eqnarray}
\Gamma_{12}^D & = & \sum \limits_n \rho_n \langle \overline{D}^0 | {\cal H}_{eff.}^{\Delta C =1} | n   \rangle
                                          \langle  n        | {\cal H}_{eff.}^{\Delta C =1} | D^0 \rangle \, ,
\\
M_{12}^D & = & \sum \limits_n \langle \overline{D}^0 | {\cal H}_{eff.}^{\Delta C =2} |D^0 \rangle 
+ P \sum \limits_n \frac{ \langle \overline{D}^0 | {\cal H}_{eff.}^{\Delta C =1} | n   \rangle
                                          \langle  n        | {\cal H}_{eff.}^{\Delta C =1} | D^0 \rangle }
                                          {m_D^2 - E_n^2} \, ,
\label{Dmix_exclusive}
\end{eqnarray}
where $n$ denotes all possible hadronic states into which both $D^0$ and $\overline{D}^0$ can decay, $\rho_n$ is the 
density of the state $n$ and $P$ is the principal value. Unfortunately, a first principle calculation 
of all the arising matrix elements is beyond our current abilities. Thus we have to make simplifying 
assumptions like only taking into account the phase space induced $SU(3)_F$ breaking effects 
and neglecting any other hadronic effects. Doing so, the authors of Refs.~\cite{Falk:2001hx,Falk:2004wg} 
found that $x$ and $y$ could naturally be of the order of a per cent. On the other hand, such a treatment 
clearly does not allow to draw strong conclusions about the existence of beyond the SM (BSM) effects.
The exclusive approach can be improved by 
using experimental input, as done in Refs.~\cite{Cheng:2010rv,Gershon:2015xra}, or by trying to take into account 
additional dynamical effects. In  Ref.~\cite{Jiang:2017zwr} the
factorization-assisted topological-amplitude approach was used for this purpose.
\subsection{Lifetimes}
The theory prediction for lifetimes of charmed hadrons relies on exactly the same theoretical framework as the 
inclusive determination of $\Gamma_{12}$ above. However, in the lifetime calculations there are no GIM cancellations present. As a result, one can gain insight whether the huge discrepancy between inclusive theory prediction
and experiment for charm mixing is due to a complete failure of the HQE or whether it is rooted in the almost perfect GIM cancellation.
 
In the charm sector we find very large ratios of lifetimes among charmed hadrons. In particular~\cite{Tanabashi:2018oca}
\begin{equation}
\left. \frac{\tau (D^+)}{\tau (D^0)} \right|_{\rm Exp.} =  \frac{(1040\pm 7) \, \rm fs}{(410.1\pm 1.5) \, \rm fs} = 2.536 \pm 0.019~.
\end{equation}
According to the HQE the lifetime of a hadron containing a heavy quark of mass $m_Q$ can be expanded as
\begin{eqnarray}
  \frac{1}{\tau} = \Gamma & = & \Gamma_0 + \frac{\Lambda^2}{m_Q^2}\Gamma_2
  + \frac{\Lambda^3}{m_Q^3}\Gamma_3
  + \frac{\Lambda^4}{m_Q^4}\Gamma_4 + ... \, .
\end{eqnarray}
The hadronic scale $\Lambda$ is of order $\Lambda^{QCD}$. Its numerical value has to be determined
by direct computation. For hadron lifetimes, $\Gamma_3$ turns out to be the dominant correction to $\Gamma_0$. 
Each of the $\Gamma_i$'s can be split up in a perturbative part and non-perturbative matrix elements. It can be
formally written as
\begin{eqnarray}
  \Gamma_i & = & \left[ \Gamma_i^{(0)} + \frac{\alpha_s}{4 \pi} \Gamma_i^{(1)}
                                      + \frac{\alpha_s^2}{(4 \pi)^2} \Gamma_i^{(2)} + ...  
    \right] \langle O^{d=i+3} \rangle  \, ,
\end{eqnarray}
where $\Gamma_i^{(0)}$ denotes the perturbative LO-contribution, $\Gamma_i^{(1)}$ the NLO one and so on;
$\langle O^{d=i+3} \rangle$ is the non-perturbative matrix element of $\Delta Q = 0$ operators
of dimension $i+3$.
The ratio $\tau (D^+) / \tau (D^0)$ is by far the theoretically best studied charm system (see  Ref.~\cite{Lenz:2018ick}) because
 $\Gamma_3^{(1)}$ and  $\Gamma_4^{(0)}$ are known \cite{Lenz:2013aua} and the hadronic matrix 
elements have been determined via 3-loop HQET sum rules \cite{Kirk:2017juj}. 
One finds   a very promising agreement with the measurement \cite{Kirk:2017juj}
\begin{equation}
\left. \frac{\tau (D^+)}{\tau (D^0)} \right|_{\rm
 HQE.} =   2.70^{+0.74}_{-0.82} = \left[1 + 16 \pi^2 (0.25)^3 (1-0.34)\right]^{+0.74}_{-0.82}\, ,
\end{equation}
indicating an expansion parameter $\Lambda/m_c \approx 0.25...0.34$, hinting at the validity of the HQE in the charm sector.
The current theory uncertainty is still dominated by the hadronic matrix elements of dimension six operators. At this point,
an independent determination with lattice QCD would be very desirable. The precision of the HQET sum rules could also be
considerably improved by performing the QCD-HQET matching at NNLO; see \textit{e.g.} Ref.~\cite{Grozin:2018wtg} for a first step in that direction.
\subsection{Duality violation in charm mixing}

The discrepancy in the HQE prediction of $\Gamma_{12}$ and the experimental value of $y$ could be resolved by including phase space dependent violations of duality of order $20\%$; see Ref.~\cite{Jubb:2016mvq}. This is another indication that there is no need for huge unknown non-perturbative effects in the charm sector.

Another interesting idea 
\cite{Georgi:1992as,Ohl:1992sr,Bigi:2000wn,Bobrowski:2010xg}
is a lifting of the severe GIM cancellation in the first and second term of Eq.~(\ref{Gamma12D})
by higher terms in the HQE. This would overcompensate for the $\Lambda/m_c$ suppression. First estimates 
of the dimension nine contribution in the HQE for $D$ mixing \cite{Bobrowski:2012jf}
indicate an enhancement compared to the leading dimension six terms. Unfortunately, this contribution is not 
large enough to explain the experimental value. A full theory determination of the HQE
terms of dimension nine and twelve will provide further insight.

It is instructive to note that the lifting of GIM cancellation in $D$-mixing by higher orders in the 
HQE \cite{Georgi:1992as,Ohl:1992sr,Bigi:2000wn,Bobrowski:2010xg} could also yield a sizeable 
CP violating phase in $\Gamma_{12}$, stemming from the second term on the r.h.s. of Eq.~(\ref{Gamma12D}).
According to Ref.~\cite{Bobrowski:2010xg}, values of up to $1 \%$ for $\phi_{12}^D$ are not yet excluded.
After settling the issues with the inclusive theory prediction for $\Gamma_{12}^D$ one could aim for
a quark level determination of $M_{12}^D$. On a very long time-scale, direct lattice calculations might
also be able to predict the SM values for $D$-mixing by building up on methods described in Ref.~\cite{Hansen:2012tf}. 

Since we found that the expansion in $1/m_c$ and $\alpha_s(m_c)$ is applicable for total inclusive quantities like lifetimes,
we now turn to the theoretical description of the exclusive quantity $\Delta A_{CP}$ with an increased confidence in the applicability of such an expansion.

\section{SM prediction for $\Delta A_{CP}$}
\subsection{Naive expectation}
\label{sec:sec31}
The amplitude of the singly Cabibbo suppressed (SCS) decay $D^0 \to \pi^+ \pi^-$ can be expressed as
\begin{equation}
      A (D^0 \to \pi^+ \pi^-) = 
                V_{cd}V_{ud}^* \left(A_{Tree} + A_{Peng.}^d \right)
              +
                V_{cs}V_{us}^* A_{Peng.}^s
              +
                V_{cb}V_{ub}^* A_{Peng.}^b \; ,
      \label{amplitude1}
      \end{equation}
where we have split  the amplitude into a tree-level amplitude $A_{Tree}$ with the CKM structure 
$ V_{cd}V_{ud}^*$ and three penguin contributions $A_{Peng.}^q$
with the internal quark $q=d,s,b$ and the CKM structure
$ V_{cq}V_{uq}^*$. All additional, more complicated, contributions like \textit{e.g.}
re-scattering effects can be put into the same scheme. 
Using the effective Hamiltonian and the unitarity of the CKM matrix we can rewrite this expression as
\cite{Lenz:2013pwa}
\begin{equation}
A  \equiv \frac{G_F}{\sqrt{2}} \lambda_d \; { T} 
      \left[ 1 + \frac{\lambda_b}{\lambda_d} { \frac{ P}{T}}\right] \; ,
\label{amplitude4}
\end{equation}
with the CKM structures $\lambda_q =  V_{cq} V_{uq}^*$. $T$ contains pure tree-level contributions,
 but also penguin topologies (P), weak exchange (E) insertions and rescattering (R) effects and $P$ consists of tree-insertion of penguin operators and
 penguin-insertions of tree level operators:
 \begin{eqnarray}
  T & =&  \sum \limits_{i = 1,2} C_i \langle Q_i^d \rangle^{T+P+E+R} 
 -  \sum \limits_{i = 1,2} C_i \langle Q_i^s \rangle^{P+R}  \, ,
\nonumber
\\
P & = & \sum \limits_{i > 3} C_i \langle Q_i^b \rangle^T 
 -  \sum \limits_{i = 1,2}    C_i \langle Q_i^s \rangle^{P+R} \, .
\end{eqnarray}
 Physical observables, like branching ratios or CP asymmetries, can be expressed in terms of $|T|$, $|P/T|$ and the strong
phase $\phi = \arg( P/T)$ as
\begin{eqnarray}
\textrm{Br} &\propto & \frac{G_F^2}{2} |\lambda_d|^2 { |T|^2}  \left|  1 + \frac{\lambda_b}{\lambda_d} { \frac{ P}{T}} \right|^2\; ,
\\
 a_{CP}^{\rm dir} & = &  \frac{-2 \left|\frac{\lambda_b}{\lambda_d}\right| \sin \gamma \left|\frac{ P}{ T}  \right| \sin \phi}
                                      {1  - 2 \left|\frac{\lambda_b}{\lambda_d}\right| \cos  \gamma  \left|\frac{ P}{ T}  \right| \cos \phi
                                        +  \left|\frac{\lambda_b}{\lambda_d}\right|^2 \left|\frac{ P}{ T}  \right| ^2}
%                 \nonumber
 %                \\
               \approx   -13 \times 10^{-4}
                 \left|\frac{ P}{ T}                \right| \sin \phi \; .
                 \label{CPasym}
\end{eqnarray}
For the $D^0 \to K^+ K^-$ decay the same formalism applies with some obvious replacements.
The branching ratios are quite well measured:
\begin{eqnarray}
\mathrm{Br}(D^0 \to K^+ K^-) & = &  (3.97 \pm 0.07) \times 10^{-3},
\nonumber
\\
\mathrm{Br}(D^0 \to \pi^+ \pi^-) & = & (1.407 \pm 0.025) \times 10^{-3},
\end{eqnarray}
and can be used to extract the size of $T$.
In the last line of Eq.~(\ref{CPasym}) numbers from the web-update of Ref.~\cite{Charles:2004jd} have been used for $\lambda_b$, $\lambda_d$ 
($|\lambda_b/\lambda_d| \approx 7 \times 10^{-4}$) and $\gamma= 65.81^\circ$. The negative sign in the CP asymmetry arises from the negative 
value of the CKM element $V_{cd}$.
Since we have $\lambda_d \approx - \lambda_s$
we expect different signs for the direct CP asymmetries in the $\pi^+\pi^-$ and $K^+ K^-$ channels.
In order to quantify the possible size of direct CP violation, we only need to know $P/T$ and
the strong phase $\phi$. One can take the naive perturbative estimate $|P/T| \approx 0.1$ \cite{Lenz:2013pwa} and get
\begin{eqnarray}
\left| a_{CP}^{\rm dir}  \right| & \leq & 1.3 \times 10^{-4} \, ,
\nonumber
\\
\left| \Delta A_{CP} \right| & \approx & 13 \times 10^{-4} \left|  \left|\frac{ P}{ T}  \right|_{K^+K^-} \sin \phi _{K^+K^-}  + \left|\frac{ P}{ T}  \right|_{\pi^+\pi^-} \sin \phi _{\pi^+\pi^-} \right|
\leq 2.6 \times 10^{-4} \,.
\end{eqnarray}
This upper bound is roughly an order of magnitude smaller than the current experimental value in Eq.~(\ref{DeltaACP_EXP}). 

 We will now discuss the LCSR calculation of $\Delta A_{CP}$ in order to determine if it is possible that non-perturbative effects can 
enhance $| P / T|$ by one order of magnitude.

\subsection{LCSR estimate}
Light-Cone Sum Rules (LCSR) \cite{Balitsky:1989ry} are a QCD based method 
allowing to determine hadronic matrix elements including non-perturbative effects. 
This method was used by the authors of Ref.~\cite{Khodjamirian:2017zdu}
to predict the CP asymmetries in the neutral $D$-meson decays.
In this paper, the values of matrix element $|T|$ were extracted from 
the experimental measurements of the branching ratios of $D^0 \to K^+ K^-$ and
$D^0 \to \pi^+ \pi^-$, and the magnitudes and phases of $P$ 
were determined using LCSRs in the same way as it was done before for non-leptonic 
$B \to \pi \pi$ decays \cite{Khodjamirian:2000mi, Khodjamirian:2003eq}.
Within this framework they get for the ratios of penguin 
to tree-level matrix elements the following values:
\begin{eqnarray}
 \left|\frac{ P}{ T}  \right|_{\pi^+\pi^-}  & = &  0.093 \pm 0.011 \, ,
\nonumber
\\
 \left|\frac{ P}{ T}  \right|_{K^+K^-}  & = & 0.075 \pm 0.015 \,.
\end{eqnarray}
It is interesting to note that these numbers agree very well with our naive estimates from the previous section.
Note that the authors of Ref.~\cite{Khodjamirian:2017zdu} do not predict the relative strong phase 
between the tree-level $T$ and penguin $P$ contributions. 
As a result, this relative phase remains a free parameter. 
Allowing for arbitrary relative strong phases yields the following bounds 
for the direct $CP$ asymmetries \cite{Khodjamirian:2017zdu}:
 \begin{eqnarray}
|a_{CP} ^{\rm dir}(D^0 \to \pi^+ \pi^-)| & \le & (1.2 \pm 0.1) \times 10^{-4}\, ,
\nonumber \\
|a_{CP} ^{\rm dir}(D^0 \to K^+ K^-)| & \le & (0.9 \pm 0.2) \times 10^{-4} \, ,
\nonumber \\
|\Delta  A_{CP}| & \le & (2.0  \pm 0.3) \times 10^{-4} \, .
\label{DeltaACP_LCSR-bounds}
\end{eqnarray}
In addition, the authors of Ref.~\cite{Khodjamirian:2017zdu} quote the following predictions: 
\begin{eqnarray}
 a_{CP}^{\rm dir} (D^0 \to \pi^+ \pi^-)& = & (-1.1 \pm 0.1) \times 10^{-4} \, ,
\nonumber \\
 a_{CP} ^{\rm dir}(D^0 \to K^+ K^-) & = & (+0.9 \pm 0.2) \times 10^{-4} \, ,
\nonumber \\
\Delta  A_{CP}            & = & (+2.0  \pm 0.3) \times 10^{-4} \, ,
\label{DeltaACP_LCSR}
\end{eqnarray}
based on the assumption of vanishing strong phases of the tree-level amplitudes $T$.
 
Because of the severe consequences of the results in Eqs.~(\ref{DeltaACP_LCSR-bounds}) 
and (\ref{DeltaACP_LCSR}) we would like to make some comments regarding these values
and to briefly investigate in what direction the work of Ref.~\cite{Khodjamirian:2017zdu} 
could be further improved. 
First, note that uncertainties quoted in Eq.~(\ref{DeltaACP_LCSR}) are pure parametric
and do not account for several missing factors discussed below.
As the authors  state, the amplitude $T$ contains matrix elements of 
different topologies, which can generate non-trivial strong phases in $T$,
as one can see in section~\ref{sec:sec31} -- 
this is neglected in the current version~\cite{Khodjamirian:2017zdu}.
Moreover, note that in the determination of $P$ the authors of
Ref.~\cite{Khodjamirian:2017zdu} neglected contributions of pure penguin operators
$Q_{i \ge 3}$ due to smallness of the corresponding Wilson coefficients.
In the analysis only contributions due to two-particle twist-2 and twist-3 
of the pion (kaon) light-cone distribution amplitudes were kept. 
It is also important to stress that they used the calculation of the 
penguin topology hadronic matrix elements of $B \to \pi \pi$ decays performed 
in Refs.~\cite{Khodjamirian:2000mi, Khodjamirian:2003eq} and adapted 
it for $D$ meson decays. But in the case of the charm meson decays 
such a computation suffers from larger uncertainties 
due to bigger values of power corrections $\sim \Lambda_{QCD} /m_c$, 
which are more sizeable compared to the case of $B$ meson decays. 

As it is pointed out in Ref.~\cite{Hansen:2012tf}, the calculation of the $D$-meson decays on the lattice is
quite challenging due to an appearance of many open channels including two, four, six, etc.
pions as well as $K\bar{K}$ and $\eta\eta$ states. In the framework of the LCSR method, contributions
of the excited states with quantum numbers of pion and $D$-meson (see definition of the
corresponding correlation function in eq. (20) in Ref.~\cite{Khodjamirian:2017zdu})
are absorbed in the spectral
density function approximated by means of quark-hadron duality that leads to introducing new effective threshold parameters
$s_0^{\pi,D}$ (see Ref.~\cite{Khodjamirian:2017zdu} for more details).

We can naively estimate the size of the higher ($>3$) twist effects ($\sim 15$ \%), 
higher perturbative radiative corrections ($\sim 13$ \%), missing terms in the OPE proportional 
to $O (s_0 / m_D^2)$ ($\sim 30$ \%) as well as systematic uncertainties related to the assumption 
of quark-hadron duality ($\sim 30$ \%) and of missing contributions of 
the penguin operators $Q_{i\ge 3}$ ($\sim 40$ \%). After adding all them in quadratures, we expect for the ratios of the matrix elements
the values with larger uncertainties:
\begin{eqnarray}
 \left|\frac{ P}{ T}  \right|_{\pi^+\pi^-}  & = &  0.093 \pm 0.056 \, ,
\nonumber
\\
 \left|\frac{ P}{ T}  \right|_{K^+K^-}  & = & 0.075 \pm 0.048 \,,
\end{eqnarray} 
which would then modify the SM bound for $\Delta A_{CP}$ to
\begin{equation}
|\Delta A_{CP}| \le (2.2 \pm 1.4) \times 10^{-4} \le 3.6 \times 10^{-4}.
\label{DeltaACP_estimate}
\end{equation}
To be conservative we will use as an upper bound the value $3.6 \times 10^{-4}$ in our BSM analysis.
Note that the central value in Eq.~(\ref{DeltaACP_estimate}) slightly differs 
from Eq.~(\ref{DeltaACP_LCSR-bounds}) due to using more recent input 
for the CKM parameters~\cite{Charles:2004jd}.

Finally, one could compute both $T$ and $P$ hadronic matrix elements entirely with the LCSR method.
In that case, one would be able to predict the relative strong phases and as 
a consequence get a more robust SM prediction for $\Delta A_{CP}$, 
instead of the estimate in Eq.~(\ref{DeltaACP_LCSR-bounds}). 
This is a time intensive calculation and we postpone it to a future study.
\section{BSM explanations of CP violation in charm decays}
\begin{figure}[t]
  \hspace{-0.7cm}\includegraphics[width=0.53\columnwidth]{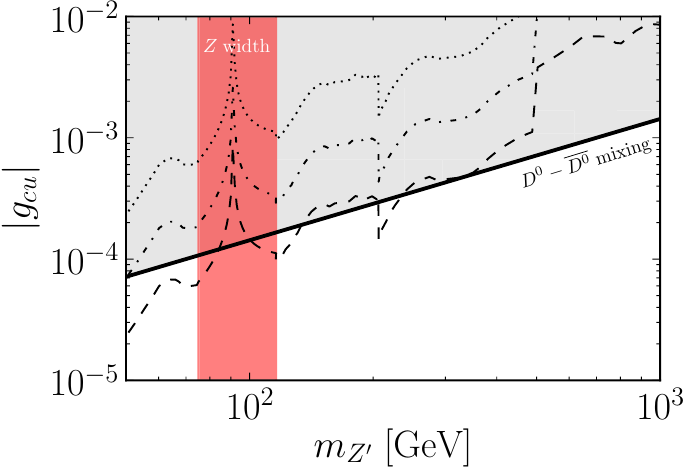}
 \includegraphics[width=0.53\columnwidth]{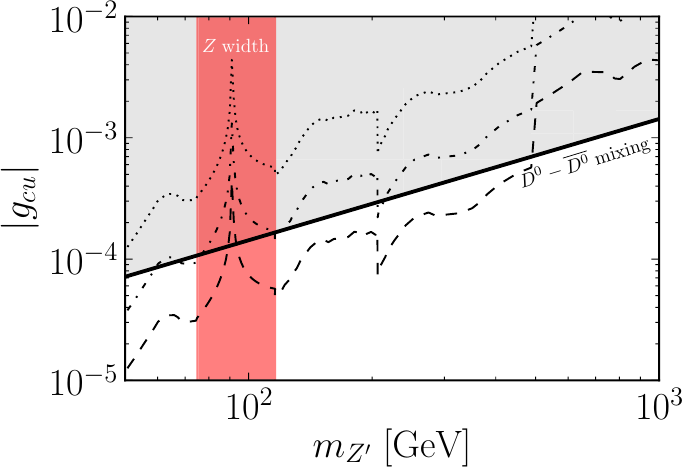}
  \caption{\it The values of $|g_{cu}|$ as a function of $m_{Z'}$ that explain  $\Delta_{NP} = -11.8\times 10^{-4}$ (left panel) and $\Delta_{NP}=-6.0\times 10^{-4}$ (right panel). The dotted line corresponds to $|\tilde{A}_{BSM}^s| = 0.1$, the dashed-dotted line stands for $|\tilde{A}_{BSM}^s| = 1/3$ and the dashed line corresponds to $|\tilde{A}_{BSM}^s| = 1$. The grey region is ruled out by $D^0-\overline{D}^0$ oscillations. We have fixed $g_{ss}$ to the maximum value allowed by collider experiments; see the text. Within the red region, the dominant constraint comes from the $Z$ width, while outside this region the dijet searches are more stringent.}\label{fig:zprime}
\end{figure}

One of the simplest explanations of the anomaly relies on extending the SM with a $Z'$ with flavour-non-diagonal couplings. The new physics contribution needs to explain the difference between the SM prediction and the experimental value. The minimum amount of asymmetry needed to reconcile the theoretical bound, Eq.~(\ref{DeltaACP_estimate}), and experimental value, Eq.~(\ref{DeltaACP_EXP}), is given by 
\begin{equation}
 \Delta_{NP} = \Delta A_{CP}^{\rm Exp.} - \Delta A_{CP}^{SM} = (-11.8\pm 2.9) \times 10^{-4}.
\end{equation}
We will assume that the relevant Lagrangian reads: 
\begin{equation}
 \mathcal{L}_{BSM} = \frac{1}{2}m_{Z'}^2 Z_\mu' Z^{'\mu}+ Z'_\mu \bigg[g_{dd} \overline{d_L}\gamma^\mu d_L + g_{ss}\overline{s_L}\gamma^\mu s_L + (g_{cu}\overline{u_L}\gamma^\mu c_L + \text{h.c.})\bigg]~.
\end{equation}
The amplitude of the $D^0 \to K^+ K^-$ decay in this case takes the form
\begin{eqnarray}\label{eq:bsm}
 A & = & \frac{G_F}{\sqrt{2}} \left( \lambda_s T + \lambda_b P \right) + \frac{1}{4} \frac{g_{cu} g_{ss}}{m_{Z'}^2} A_{BSM}^s
 =
 \frac{G_F}{\sqrt{2}}   \lambda_s T \left[ 1+ \frac{\lambda_b}{\lambda_s}  \frac{P}{T} + \tilde{g}^2_s \tilde{A}_{BSM}^s \right] \, ,
 \end{eqnarray}
 where
 \begin{align}
 \tilde{g}^2_s \equiv  \frac{ \sqrt{2} g_{cu} g_{ss}}{4 G_F \lambda_s m_{Z'}^2}~, \quad  \tilde{A}_{BSM}^s \equiv  \frac{A_{BSM}^s}{T} 
 = \frac{\langle K^+ K^- | \bar{u} \gamma^\mu (1-\gamma_5) c \; \bar{s} \gamma^\mu (1-\gamma_5) s | D^0\rangle }
           {\langle K^+ K^- | \bar{u} \gamma^\mu (1-\gamma_5) s \; \bar{s} \gamma^\mu (1-\gamma_5) c | D^0\rangle}~,
 \end{align}
 where the last equality is to leading order in $\alpha_s$. Naive colour counting yields $ |\tilde{A}_{BSM}^s| \approx 1/N_c$; to be conservative we will use below $ |\tilde{A}_{BSM}^s| \in [0.1, 1]$.
From here on, we neglect the SM penguin-tree level ratio, because that is the main source of the SM contribution, $\Delta A_{CP}^{SM}$ mentioned before. This implies that the new physics contribution to the direct CP asymmetry reads:
 \begin{eqnarray}
 a_{CP}^{\rm dir} & = &  \frac{2 \left| \tilde{g} _s\right|^2  \left| \tilde{A}_{BSM}^s\right| \sin \delta_{BSM}^s   \sin \phi_{BSM}^s}
                                        {1  - 2  \left| \tilde{g_s} \right|^2  \left| \tilde{A}_{BSM}^s \right| \cos \delta_{BSM}^s  \cos \phi_{BSM}^s+ \left| \tilde{g}_s \right|^4  \left| \tilde{A}_{BSM}^s \right|^2 }
                                        \nonumber
                                        \\
                                        &\approx& 2 \left| \tilde{g} _s\right|^2  \left| \tilde{A}_{BSM}^s \right| \sin \delta_{BSM}^s  \sin \phi_{BSM}^s\, ,
\end{eqnarray}
with $ \delta_{BSM}^s= \arg     ( \tilde{g}_s^2 )$ and          $ \phi_{BSM}^s= \arg     ( \tilde{A}_{BSM}^s )$.     The generalisation to the $\pi^+ \pi^-$
case is straightforward. 
To explain the central value of $\Delta_{NP}$ within this model we need
\begin{eqnarray}                
   \Delta_{NP}  &= & 2 \left| \tilde{g} _s\right|^2  \left| \tilde{A}_{BSM}^s \right| \sin \delta_{BSM}^s  \sin \phi_{BSM}^s
   \nonumber \\ &&
                                                         - 2 \left| \tilde{g} _d\right|^2  \left| \tilde{A}_{BSM}^d \right| \sin \delta_{BSM}^d  \sin \phi_{BSM}^d~.
\end{eqnarray}
Let us assume, for now, that the whole effect originates in the $K^+ K^-$ final state, namely $g_{dd} = 0$. We get:
    \begin{eqnarray}                
   \Delta_{NP}&= &   2 \left| \tilde{g} _s\right|^2  \left| \tilde{A}_{BSM}^s \right| \sin \delta_{BSM}^s  \sin \phi_{BSM}^s\,
   \nonumber
    \\
    \Rightarrow 
    |g_{cu}| &=& \Delta_{NP} \frac{\sqrt{2} G_F \lambda_s m_{Z'}^2}{ g_{ss}} \left(  \left|  \frac{A_{BSM}^s}{T} \right| \sin \delta_{BSM}^s  \sin \phi_{BSM}^s\right)^{-1}.
\end{eqnarray}
 Fixing $\sin{\delta_{BSM}^s}\sin{ \phi_{BSM}^s} = -1$, we plot in Fig.~\ref{fig:zprime} the value of $|g_{cu}|$ as a function of $m_{Z'}$ for different choices of $|\tilde{A}_{BSM}^s|$, for a central value of $\Delta_{NP} = -11.8\times 10^{-4}$ (left panel) and for a two-sigma departure of $\Delta_{NP}=-6.0\times 10^{-4}$ (right panel).

We have fixed $g_{ss}$ to the maximum value allowed by the most stringent LHC constraints, provided by the CMS analysis of Ref.~\cite{Sirunyan:2017nvi} 
and by the constraint on the width of the SM $Z$~boson. 
By tagging $Z'$ production with an additional initial state radiated jet, the CMS search explores masses as small as $50$ GeV, superseding previous searches by UA2 and CDF.
In order to estimate the upper bound on $g_{ss}$ from dijet searches, we computed the $Z'$ production cross section in $pp$ collisions at $\sqrt{s} = 13$ TeV for masses in the range $m_{Z'}\in [50,1000]$ GeV, using \texttt{MadGraph}~\cite{Alwall:2014hca} with an \texttt{UFO} model~\cite{Degrande:2011ua} implemented in \texttt{Feynrules}~\cite{Alloul:2013bka}. 
Denoting by $\sigma_{\text{theory}}$ the theoretical cross-section for $g_{ss} = 1$ and  $\sigma_{\text{limit}}$ the experimental limit provided in the left panel of Fig. 7 of Ref.~\cite{Sirunyan:2017nvi}, we obtain
\begin{equation}
g_{ss}^\text{max} = \sqrt{\frac{\sigma_{\text{limit}}}{\sigma_{\text{theory}}}}~.
\end{equation}

Loops of strange quarks induce mixing between the $Z$ and the $Z'$, which in turn leads to corrections to the width of the $Z$. This width is well measured~\cite{Tanabashi:2018oca} and implies an upper bound on $g_{ss}$. In order to estimate the minimal allowed mixing between the $Z$ and the $Z'$ (without fine tuning) we set this mixing to zero at a cut-off scale $\Lambda$, $c_Z(\Lambda) = 0$. The RG flow of the mixing parameter is governed by~\cite{Graesser:2011vj}:
\begin{equation}
    \mu\frac{d c_Z}{d \mu} = - \frac{g_{ss}e}{32\pi^2 s_w^2 c_w} \left[ 3 -4 s_w^2\right]~,
\end{equation}
with $e = \sqrt{4\pi\alpha}$ and $s_w$ and $c_w$ the sine and cosine of the Weinberg angle.
The $Z-Z'$ mixing is  of order $c_Z(m_Z) = 0.024 g_{ss}$ for $\Lambda = \mathcal{O}(1)$ TeV. This in turn introduces a correction to the width of the $Z$ boson~\cite{Dobrescu:2014fca}:
\begin{equation}
    \frac{\Delta \Gamma_Z}{\Gamma_Z} = \frac{g_{ss} c_Z s_w c_w V_d}{3g(1-M_{Z'}^2/M_Z^2)(2V_u^2 + 3V_d^2 + 5/16)},
\end{equation}
where $V_{u,d} = \pm 1/4 - (3\pm 1)s_w^2/6$. This translates to a bound on $g_{ss}$, which we show in Fig.~\ref{fig:zprime}.

The $g_{uc}$ coupling induces $D^0-\overline{D}^0$ mixing. We take the limit from Ref.~\cite{Altmannshofer:2012ur} and show it also in the Fig.~\ref{fig:zprime}. At first, it may seem that since the experimental value of $x$ has changed by a factor of two since the analysis of Ref.~\cite{Altmannshofer:2012ur}, this analysis may be no longer applicable in its original form. However, the relevant change in the range of possible long-distance contribution to the $M_{12}$ due to change in $x$ is only $10\%$. In order to repeat the method of  Ref.~\cite{Altmannshofer:2012ur} we would need to scan over a range $M_{12}^{LD} \in [-0.017\, \mbox{ps}^{-1},0.017 \,\mbox{ps}^{-1}] $  and $\Gamma_{12}^{LD} \in [-0.036\,\mbox{ps}^{-1},0.036\,\mbox{ps}^{-1}]$, as compared to previous ranges of
$M_{12}^{LD} \in [-0.02\,\mbox{ps}^{-1},0.02\,\mbox{ps}^{-1}]$ and $\Gamma_{12}^{LD} \in [-0.04\,\mbox{ps}^{-1},0.04\,\mbox{ps}^{-1}]$. For this reason, we simply adopt the previous results.

We note that the central value of the LHCb measurement can be explained within this model provided $m_{Z'}\lesssim 80$ GeV. It may seem possible to avoid the dijet bounds by allowing the $Z'$ to decay into other final states. However, this is not the case. The cross-sections for $Z'$ decaying into two particle final states such as light leptons, taus and bottoms are constrained to be a factor of $\sim 3000$, $\sim 100$ and $\sim 30$ smaller than the dijet cross-section~\cite{ATLAS-CONF-2019-001,Khachatryan:2016qkc,Sirunyan:2018ikr}. On other hand, invisible decays of the $Z'$ are severely constrained by the monojet searches as shown in Ref~\cite{Chala:2015ama}.
The fact that we can resolve  $\Delta A_{CP}^{\rm Exp.}$ for $m_{Z'} < 80$ GeV motivates further searches for light $Z'$ bosons.

Had we assumed that the anomaly is due mostly to the $\pi^+\pi^-$ decay, the production cross-section for $Z'$ would be enhanced by the larger $d$-quark parton distribution functions by a factor of $x \sim 4$. This would increase the $g_{uc}$ necessary to explain the $\Delta A_{CP}^{\rm Exp.}$ by a factor of $\sqrt{x} \sim 2$, effectively ruling out most of the parameter space of the model. In principle, it is possible to arrange for the new physics to contribute to $\Delta A_{CP}$ from both $K^+K^-$ and $\pi^+\pi^-$ decays, but the gain in such scenario is minimal.

We would like to highlight the fact that $Z'$ models have been suggested as solution to other flavour anomalies such as violation of lepton universality in $B \to K^* l^+l^-$  \cite{Altmannshofer:2013foa}. Whether these mechanisms can be unified in a single framework is a good starting point for future work.

Finally, let us comment on other simple extensions of the SM that could explain the measured value of $\Delta A_{CP}$. These include a $W'$ and a heavy gluon $G$. The former can not introduce a new source of CP violation, because it involves an identical strong matrix element and therefore $\sin\delta^s_{BSM} = 0$.
A class of quirky solutions may come from arranging for destructive interference between the SM tree-level and new physics contributions in the Kaon final state decays. This would lead to a significant enhancement of the contribution of the penguin diagrams to the CP violation and could produce large $\Delta A_{CP}^{\rm Exp.}$. Unfortunately, such a change of the matrix element leads to a significant change of the partial decay width $D^0 \rightarrow K^+K^-$ and appears unfeasible.
Regarding the heavy gluon, masses above $100$ GeV are excluded irrespectively of $g_{ss}$ or $g_{dd}$, since they can be pair produced in a model-independent way via QCD and no significant excess over the SM background has been observed in the corresponding searches; see Refs.~\cite{Khachatryan:2014lpa,Aaboud:2017nmi}. To the best of our knowledge, there are no searches for pair-produced massive gluons with mass below $100$ GeV.

\section{Conclusion}
Compared to the situation in 2011, we have learnt that HQE tools 
can successfully describe the lifetime ratio of charmed mesons. 
The apparent failure of the HQE for $D$ mixing, the naive estimate of a
correction of the order $10^4$ might come from a non-perturbative effect as small as $20\%$.
These new theory developments increase our confidence in first principle QCD methods,
like LCSR, for the charm sector. Within this framework we find a maximal value of 
$|\Delta A_{CP}^{\rm SM}| \leq 3.6 \times 10^{-4}$, which deviates significantly 
from the experimental result.
The next steps to further strengthen our confidence in the theory tools would be a higher theoretical
precision in the lifetime predictions of $\tau (D^+) / \tau(D^0)$ due to a determination of the
arising non-perturbative matrix elements with lattice QCD - a higher precision can also be obtained 
within the HQET sum rule approach if the HQET-QCD matching will be performed at NNLO. 
Next these calculations (lattice and HQET sum rules) should be extended to lifetimes 
ratios of the $D_s^+$ meson and charmed baryons, where we have so far only LO estimates 
\cite{Cheng:2018rkz}, that seem to be affected by cancellations peculiar to the exclusive use
of LO expressions. 
For a full confidence on claiming a BSM origin of the measured $\Delta A_{CP}$ value, also a first
principle determination of the tree-level contributions to the decays $D \to \pi \pi$ 
and $D \to K K$ will be necessary.
One can in principle compute both tree-level and penguin hadronic matrix elements 
entirely within the Light-Cone Sum Rules method (following Ref.~\cite{Khodjamirian:2017zdu}), 
despite accounting of different topologies in the tree-level matrix element within 
this method will require much more computing efforts.
This will allow to determine not only the magnitudes of both matrix elements and their ratio
but also to predict relative strong phase.
As a consequence one will get a real SM prediction for $\Delta A_{CP}$ (not just a bound for magnitude)
from the first principles of QCD providing an additional test of the Standard Model in the charm sector.

Thus we have also explored the possibility of explaining this discrepancy by extending the SM with a leptophobic $Z'$ with flavour-violating couplings to $\overline{c}u$ quarks and flavour-conserving couplings to $\overline{s}s$ quarks, without conflicting with dijet searches at colliders, measurements of the SM $Z$-boson width and $D^0-\overline{D}^0$ oscillation data. We show that this is feasible for $m_{Z'}\lesssim 80$ GeV and $|g_{uc}|\sim 10^{-4}$ for maximal value of $\tilde{A}_{BSM}^s$. For the most likely value of  $\tilde{A}_{BSM}^s \sim 1/3$, one can still explain the anomaly provided $m_{Z'} \lesssim 60$ GeV.
It is exciting that off-diagonal couplings in the down sector of the same order of magnitude can address the $R_{K^{(*)}}$ anomalies as well~\cite{DiLuzio:2017fdq}.
There are no constraints from dijet searches below $m_{Z'} \sim 50$ GeV, and so this anomaly motivates further experimental effort in the low mass $Z'$ frontier.

\section*{Acknowledgments}
We would like to thank the participants of the TUPIFP workshop in Durham 
for many interesting and useful discussions, in particular Hai-Yang Cheng, 
Yuval Grossman, Alexander Khodjamirian and Fu-Sheng Yu.
This works was supported by STFC via the IPPP grant. MC is funded by the Royal Society under the Newton International Fellowship programme.

\section*{Note added} After this work was finished two papers appeared on the arXiv \cite{Grossman:2019xcj,Li:2019hho}
attributing the new LHCb measurement of $\Delta A_{CP}$ to SM effects. We acknowledge the line of thought in these two papers,
but we do not see the necessity for assuming an $\mathcal{O}(10)$ enhancement of hadronic effects over the perturbative SM estimate, even if this assumption is self-consistent.

\bibliographystyle{elsarticle-num}
\bibliography{references}

\end{document}